\documentclass[useAMS,usenatbib]{mn2e}

\usepackage{amsmath,amssymb}
\usepackage{graphicx}
\usepackage{lineno}
\setcitestyle{authoryear}

\DeclareGraphicsExtensions{.pdf,.png,.jpg,.eps}
\bibpunct{(}{)}{;}{a}{}{,} 

\title[MAGIC detection of the low-luminosity blazar 1ES\,1741+196]
{MAGIC detection of very high energy $\gamma$-ray emission from the low-luminosity blazar 1ES\,1741+196}

\author[M.~L.~Ahnen~et.~al.]{
M.~L.~Ahnen$^{1}$,
S.~Ansoldi$^{2,24}$,
L.~A.~Antonelli$^{3}$,
P.~Antoranz$^{4}$,
C.~Arcaro$^{5}$,\newauthor 
A.~Babic$^{6}$,
B.~Banerjee$^{7}$,
P.~Bangale$^{8}$,
U.~Barres de Almeida$^{8,25}$
% ...
\\
(Affiliations can be found after the references)}

\begin{document}

%\date{Submitted \today}

\maketitle

%{*} Corresponding authors: Saverio Lombardi (saverio.lombardi@oa-roma.inaf.it)\\, Nijil Mankuzhiyil (mankuzhiyil.nijil@gmail.com), Michele Palatiello (michele\\.palatiello@gmail.com), 
%Massimo Persic (massimo.persic@gmail.com), Malwina Uellenbeck (malwina.uellenb\\eck@tu-dortmund.de), 
%and Sara Buson (sara.buson@pd.infn.it)

\date{Received: ???~/~Accepted: ???}

\begin {abstract} 
{
 We present the first detection of the nearby (z=0.084) low-luminosity BL Lac object 1ES\,1741+196 in the very high energy (VHE:\,E$>$100\,GeV) band. This object lies in a triplet of interacting galaxies. Early predictions had suggested 1ES\,1741+196 to be, along with several other high-frequency BL\,Lac sources, within the reach of MAGIC detectability.  Its detection by MAGIC, later confirmed by VERITAS, helps to expand the small population of known TeV BL\,Lacs. The  source was observed with the MAGIC telescopes between   2010 April and 2011  May, collecting 46\,h of good quality data.  These observations led to the detection of the source at 6.0\,$\sigma$ confidence level, with a steady flux $\mathrm{F}(> 100\,{\rm GeV}) = (6.4 \pm 1.7_{\mathrm{stat}}\pm 2.6_{\mathrm{syst}}) \cdot 10^{-12}$\, ph\,cm$^{-2}$s$^{-1}$ and a differential spectral photon index $\Gamma = 2.4 \pm 0.2_{\mathrm{stat}} \pm 0.2_{\mathrm{syst}}$ in the range of {$\sim$80\,GeV - 3\,TeV}. To study the broad-band spectral energy distribution (SED) simultaneous with MAGIC  observations, we use KVA, {\it Swift}/UVOT and XRT, and {\it Fermi}/LAT data. One-zone synchrotron-self-Compton (SSC) modeling of the SED of 1ES\,1741+196 suggests values for the SSC parameters that are quite common among known TeV BL\,Lacs  except for a relatively low Doppler factor and slope of electron energy distribution. A thermal feature seen in the SED is well matched by a giant elliptical's template. This appears to be the signature of thermal emission from the host galaxy, which is clearly resolved in optical observations.
}
\end{abstract}

\begin {keywords}
galaxies: galaxies: BL Lacertae objects: individual (1ES\,1741+196) - gamma-rays: galaxies.
\end{keywords}

%\titlerunning{Discovery of 1ES\,0033+595 at very high energy by the MAGIC telescopes}

%\titlerunning{1ES\,0033+595 VHE discovery by MAGIC}

%\authorrunning{Aleksi\'c et~al. (MAGIC collaboration)}

%\maketitle

\section{Introduction}

 Blazars are thought to be black-hole (BH) powered Active Galactic Nuclei (AGN) whose relativistic jets are closely aligned with our line-of-sight. They constitute the most numerous class of detected extragalactic very-high-energy (VHE: E$>$100\,GeV) $\gamma$-ray sources. Their spectral energy distribution (SED) typically shows two emission components: {\it (i)} one component peaks at eV-keV energies, interpreted as synchrotron radiation emitted by relativistic electrons moving in the jet's magnetic field; and {\it (ii)} another component, which peaks at $\gamma$-ray frequencies, commonly interpreted as  arising from inverse Compton (IC) scattering of lower-energy photons (\citealp{rees67}) -- the latter being either the above-mentioned synchrotron photons internal to the jet (Synchro-Self-Compton (SSC) scenario, see Maraschi, Ghisellini \& Celotti 1992) or some other photon field external to the jet (External Compton (EC) scenario, see \citealp{der_schlick93}). The high energy peak  may also result from hadronic processes, as proposed by Mannheim (1993).   BL Lac objects are blazars characterized by a featureless, highly polarized, broad-band (radio to VHE) continuum emission. %that shows rapid variability at all frequencies and on all time scales probed so far.

1ES\,1741+196 is   a high-frequency-peaked BL\,Lac object (HBL; where the synchrotron peaks at X-ray, while its IC counterpart peaks at VHE), at  coordinates (J2000) RA = 17:43:57.8 (hh:mm:ss) and DEC=19:35:09 (dd:mm:ss), at redshift $z=0.084$. Its host galaxy is one of the most luminous and largest ($M_R$=-24.85; $r_e$=51.2\,kpc) among  BL Lac host galaxies. Two nearby (7.2 and 25.2 kpc) companion galaxies at the same redshift suggest that 1ES\,1741+196 could be a BL Lac object in a triplet of interacting galaxies (\citealp{heidt1999}).  It was detected in radio, optical, X-ray and high energy (HE: E$>$100\,MeV) $\gamma$-ray frequencies (\citealp{radio1741}; \citealp{heidt1999}; \citealp{xray1741}; \citealp{fermi3}). Its high resolution radio map \citep{radiomap} shows a parsec scale one-sided jet. The jet-counterjet brightness ratio suggests a Doppler factor of $\delta>$4, for a viewing angle of a few degrees. 

  Prompted by the prediction of TeV flux based on the BeppoSAX observations and the SSC model \citep{cos_ghis02}, MAGIC observed this source in mono-mode for a total of 16\,hr between 2007\,July and 2008\,August, obtaining a significance of $2\sigma$, and a flux upper limit of F($>$170\, GeV)$ < 3.6 \cdot 10^{-11}$ph\,cm$^{-2}$ s$^{-1}$ \citep {magicupperlimit}. Further MAGIC observations carried out between  2010\,April and 2011\,May in stereoscopic mode finally led to the detection of the source at VHE\,$\gamma$-ray frequencies (\citealp{berger11}).  This was later confirmed by VERITAS (\citealp{veritas}).

 In this paper we study the emission features of the 1ES\,1741+196, the only BL Lac object detected in  a triplet of interacting galaxies, using the data collected from MAGIC and other multi-frequency instruments.
 In Section\,2 we describe the multi-frequency data used for this analysis. The results are  presented in Section\,3, discussed in Section\,4, and summarized in Section\,5.

\section{Observations and Data Analysis}

Observations of 1ES\,1741+196 during time periods that include the MAGIC observation window  were performed in the optical, X-ray and HE\,$\gamma$-ray ranges, which are discussed in detail in the following sections.

\subsection{KVA}

The KVA (Kungliga Vetenskapsakademien Academy)\footnote{Tuorla Blazar monitoring program, 
http://users.utu.fi/kani} telescopes are located at La Palma but operated remotely by the Tuorla Observatory in Finland. These telescopes are used mainly for optical support observations for the MAGIC telescopes. The KVA telescopes consist of a 60\,cm telescope which is used for polarimetric observations  and a 35\,cm telescope used for photometry simultaneous with MAGIC observations. Furthermore, the smaller 35\,cm telescope 
monitors potential VHE $\gamma$-ray candidate AGNs in order to trigger MAGIC observations if one of these selected objects is in a high optical state. { These} observations are performed in the $R$-band and the magnitude of the source is measured from CCD images using differential photometry,  i.e. by comparing the brightness of the object with that of several calibrated stars in the same field of view. The data were processed by the reduction  programmes developed in Tuorla Observatory (see \citealp{Nilsson2016} and the references therein).

\subsection{{Swift}}
The {\it Swift} satellite, which was launched in 2004 \citep{Gehrels2004}  carries three instruments, the Burst Alert Telescope (BAT; sensitive 15-150 keV; \citealp{Barthelmy2005}), the X-ray telescope (XRT; sensitive 0.2-10 keV; \citealp{Burrows2005}) and the UV/Optical Telescope (UVOT; sensitive 170-600 nm; \citealp{Roming2005}). 

The {\it Swift}/XRT data which fall in the MAGIC observation period were taken  in Photon Counting mode  on 2010 July 30 and 2011 January 21.  These data were  processed  by the  XRTPIPELINE (version 0.13.1) distributed by HEASARC within  the  HEASoft  package (v.6.16) using standard procedure. Events with grades 0-12 were selected  (see Burrows et al. 2005) and the response matrices available in  the Swift CALDB (20110101v014) were used. The source events in the 0.3-10 keV range within a circle with a radius of 22 arcsec were selected for the spectral analysis. The background was 
extracted from off-source circular  regions of  the same radius. The spectra were extracted from the corresponding event files and binned using GRPPHA to ensure a minimum of 20 counts per energy bin. Spectral analysis was performed using XSPEC version 12.8.2.

{\it Swift}/UVOT source counts were extracted from a circular region of radius 5 arcsec, centered on the source position. The background was estimated from three circular  source free regions of the same radius. These data were processed with the uvotmaghist task of the HEASOFT package. 

\subsection{{Fermi}/LAT}

The pair-conversion Large Area Telescope (LAT) on board the {\it Fermi} satellite monitors the $\gamma$-ray sky in  survey mode every 3 hours in the energy range from 20 MeV to $>$ 300 GeV \citep{atwood09}.  The data presented in this paper were selected from a  $15^{\circ}$ radius region of interest (ROI) centered at the location of the 1ES\,1741+196, during the first 6.7 years of the mission from 2008 August 4 to 2015 April 7 (MJD 54682.7--57119.3). We analyzed the data in the energy range from 100\,MeV to 300\,GeV. %and {\bf we have searched for any hint of variability.} 
The analysis was performed with the ScienceTools software package version v9r33p0 and the instrument response function P7REP\_SOURCE\_V15 \citep{fermianalysis}.  The event selection was based on Pass\,7 reprocessed source class events and a zenith angle cut of $< 100^{\circ}$ was applied to reduce the contamination from the Earth limb.   The Galactic diffuse emission model \citep{bgmodel} and isotropic component used were gll\_iem\_v05\_rev1.fit and iso\_source\_v05.txt, respectively as recommended for Pass 7 Reprocessed Source event class\footnote{http://fermi.gsfc.nasa.gov/ssc/data/access/lat/BackgroundModels.html}. The normalizations of both components in the background model were allowed to vary freely during the spectral fitting. A binned maximum-likelihood method  analysis was used  \citep{mattox1996}.

For a first likelihood fit making use of gtlike, the model includes all the sources within 20$^{\circ}$  of the source of interest which are included in  the {\it Fermi}/LAT third source catalog \citep{fermi3}. For the spectral fit  (simple power law) spectral  indices and fluxes were left free for the sources within 15$^{\circ}$, while sources from 15$^{\circ}$ to 20$^{\circ}$ were frozen to the catalog value.  From the  residual of the model (created using gtmodel) with respect to the data within the ROI, in addition to the 3FGL sources, we  identified one new source with test  statistic TS=26.1 located 9.9$^{\circ}$ from 1ES 1741+196.  This was included in the model. In addition, 5 more sources with TS between 5.3 and 12.8 and located  between 9.1 and 11.9$^{\circ}$ from 1ES 1741+196 were included in the model.   The best location of these 6 additional sources were found using gtfindsrc. 
The sources with TS $<5$ were deleted from the model. A second maximum-likelihood analysis was performed on the updated source model.   For the light curve calculation in 1 year time bins shown in Fig. 2, only the source of interest and the diffuse models were left free to vary while the rest of the sources considered in the analysis were fixed to the values obtained from the analysis of the entire data sample. Also variability on monthly timescales was investigated.

\subsection{MAGIC}

MAGIC is a system of two 17\,m dish Imaging Atmospheric Cherenkov Telescopes (IACT) located at the Roque de los Muchachos observatory ($28.8^\circ$N, $17.8^\circ$W, 2200\,m a.s.l.), on the Canary Island of La Palma, Spain. Since 2009 the MAGIC telescopes operate stereoscopically, with a sensitivity of $<$0.7$\%$  crab-unit (integrated flux from the Crab Nebula) for energies $>$\,220\,GeV in 50\,hr of observations  (\citealp{aleksic+11}). 

The MAGIC telescopes observed 1ES\,1741+196 for 53 nights from 2010 April 10 until 2011 May 26, for a total observation time of approximately 57\,hr in the so-called wobble mode \citep{fomin+94}. The data were taken for zenith angles in the range from $9^{\circ}$ to $38^{\circ}$,  which resulted in an energy threshold (defined as the peak of the Monte Carlo (MC) simulated photon energy distribution for a Crab-Nebula-like spectrum after all analysis cuts) of 90\,GeV. 

After the application of standard quality checks based on the rate of the stereo events and the distributions of basic image parameters, $\sim$46\,hr of effective on-time { data} were selected. 
%The rejected data were affected mainly by non-optimal atmospheric conditions during the data taking.
\renewcommand*{\thefootnote}{\fnsymbol{footnote}}
Data analysis was performed using the standard software package MARS (\citealp{albert+08a, aliu+09}), including the latest routines for stereoscopic analysis (\citealp{aleksic+12, lombardi+11, zanin+13}). After the calibration (\citealp{albert+08b}) and the image cleaning of the events recorded by each telescope, the information coming from the individual telescopes is combined and the calculation of basic stereo image parameters is performed.  For $\gamma$/hadron separation and the $\gamma$-direction estimation, a multivariate method called Random Forest (\citealp{albert+08c}) was applied using image parameters (\citealp{hillas85}), timing information (\citealp{aliu+09}), and stereo parameters (\citealp{aleksic+12}), to compute a $\gamma$/hadron discriminator, called {\it hadronness}. 
  While computing the significance of the signal coming from the 1ES\,1741+196 sky region, we applied single cuts in {\it hadronness} and  $\theta^2$\footnote{The parameter $\theta^2$ is the squared angular distance between the reconstructed source position of the events and the nominal position of the expected source.}, which were optimised to maximise the significance  of the signal (above 250 GeV) in a Crab nebula data set.  Conversely, in deriving the spectrum and the light curve of the source, we applied different cut values in {\textit {hadronness}}   that, for each logarithmic energy bin, yield a gamma efficiency of 90\% in the MC gamma dataset. These procedures, which are regularly used to analyse MAGIC data, are described in detail in Aleksi{\'c} (2012).

%While computing the significance of the signal coming from the 1ES\,1741+196 sky region, single cuts in {\it hadronness} and ${\mathrm{\theta^2}}$\footnote{The parameter $\theta^2$ is the squared angular distance between the reconstructed source position of the events and the nominal position of the expected source.} {\bf optimized (using Crab Nebula data) for energies above $\sim 250$ GeV were applied. Conversely, in deriving the spectrum and the light curve of the source, we used different cut values optimized using MC-simulated and Off (sky region where no $\gamma$-ray emitter is present) data, in each logarithmic energy  bin.}

\renewcommand*{\thefootnote}{\arabic{footnote}}
%is applied. For the this task, the algorithm employs basic image parameters (\citealp{hillas85}), timing information (\citealp{aliu+09}), and stereo parameters (\citealp{aleksic+12}) to compute a $\gamma$/hadron discriminator, called {\it hadronness}, by comparing real (hadronic-dominated) data with MC $\gamma$-ray simulations. 
%The {\it hadronness} parameter ranges from 0 (for showers that were confidently initiated by $\gamma$-rays) to 1  (for showers that were clearly initiated by hadronic cosmic rays). Finally, the estimation of the energy of the events is achieved by averaging individual energy estimators for both telescopes based on look-up tables (\citealp{aleksic+12}). 

%The final analysis cuts applied to the 1ES\,1741+196 data were optimized by means of contemporaneous Crab Nebula data and MC simulations. In computing the significance of the signal coming from the 1ES\,1741+196 sky region, single cuts in {\it  hadronness} and $\theta^2$ optimized for energies close to the threshold were applied. Conversely, in deriving the spectrum and the light curve of the source, we used multiple cuts optimized in logarithmic energy bins.

\section{Results}

In the following sections the analysis results from the optical, X-ray, HE, and VHE data are presented.

\subsection{KVA}

\begin{figure}

%\centering
\vspace{-2.5cm}
\hspace{-1cm}
\includegraphics[width=0.65\textwidth]{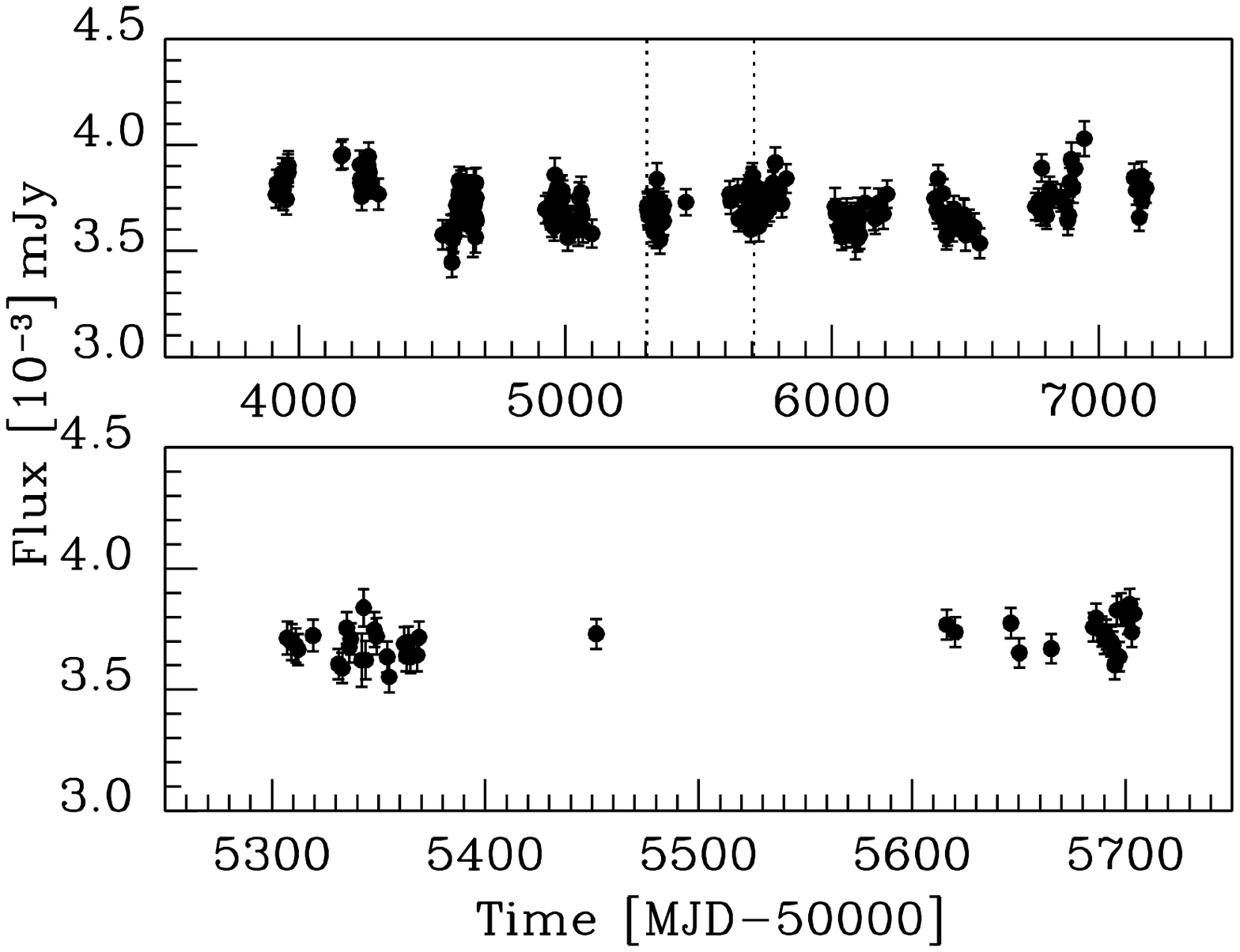}
\vspace{-7cm}
\caption
{
Optical $R$-band light curve from 7 years  of monitoring observations performed by the Tuorla Observatory. The contribution of the host galaxy has not been subtracted. The MAGIC observation window in 2010-2011 is indicated by the vertical lines in the top panel whereas the bottom panel presents the zoomed light curve in the MAGIC observation period.
}
\label{fig:1741_optical_lc}

\end{figure}

The top panel of Fig.~\ref{fig:1741_optical_lc}, shows the light curve obtained from the photometric observations of KVA, between 2006 June and 2013 November, while the bottom panel shows the light curve from 2010 April 21 to 2011 May 23  that coincides with the MAGIC observation window.

 The brightness was corrected for  the dust in the Galaxy (\citealp{sch98}).  Since the host galaxy belongs to a triplet of interacting galaxies, we used a slightly different approach to estimate the host galaxy magnitude, using a  Nordic Optical Telescope (NOT) image (see also Nilsson et al. 2007). After subtracting the central point source based on the scaling from the model fit and  convolving the image with a seeing of 2 arcsec (typical KVA good seeing value), the counts inside a radius of 7.5 arcsec (which is the KVA measurement aperture) were measured. In order to match this NOT image to the same calibration used in the case of KVA monitoring, the brightness of two unsaturated stars in the NOT image (convolved with 2 arcsec seeing) were calculated. In order to derive the transformation from counts to magnitude, we measured  the same stars (as in the NOT image) using the KVA data with the same reference star that is used in the monitoring. Using this approach the host magnitude has been computed. The flux of the host galaxy turned out to be 2.5$\pm$0.3\,mJy.

 After subtracting the estimated host galaxy flux from the observed KVA flux of 1ES\,1741+196 (averaged over the MAGIC observation; Fig.\,1), the residual KVA flux -- an average value attributable to the  nuclear region of blazar over the MAGIC observation time -- is $(1.06\pm 0.01)$\,mJy. This matches the emission level monitored over the whole 7-year span of KVA observations, i.e., $(1.07\pm 0.01)$\,mJy.
%The average $\nu F_{\nu}$, during the MAGIC observations is ($5.0\pm0.3$) $\cdot 10^{-12}$\,erg\,cm$^{-2}$\,s$^{-1}$. 
The source shows only marginal variability during the 7\,yr KVA survey (as also shown in Lindfors et al. 2016), and hardly any variability during the MAGIC observations (respectively, top and bottom panels of Fig.~\ref{fig:1741_optical_lc}). 

\subsection{{Swift}}

 The  X-ray  spectra can be well described by a  simple power law  ($\chi^{2}/{\rm d.o.f} = 1.1$) in the range of 0.3 - 10 keV, with a photon index $\mathrm{\Gamma = 1.9 \pm 0.1}$,
and a normalization constant $\mathrm{f_0} = (2.8 \pm 0.1) \cdot 10^{-3}$ keV$^{-1}$\,cm$^{-2}$\,s$^{-1}$ at 1\,keV.  The  neutral  hydrogen-equivalent column  density  was  fixed  to the Galactic value in the direction of the source, which is $7.36\times10^{20}$ cm$^{-2}$ \citep{Kalberla2005}. We have also found that there is no significant spectral variability in the XRT observations of 2010 July 30 and 2011 January 21.  %The spectral points obtained from the simple power law fit are shown in Fig.\,7.

The fluxes obtained from {\it Swift}/UVOT analysis have been corrected for Galactic extinction ${\rm E\,}_{{\rm B-V}}=0.079$ mag \citep{sch98}. The exact amount of the host galaxy contribution is not given in the literature.  Hence, we estimated the host galaxy magnitude V=1.1\,mJy, B=0.5\,mJy and U=0.1\,mJy based on the R-band value (aperture 5\,arcsec) from Nilsson et al. (2007) by using galaxy colours at z=0  (Fukugita, Shimasaku \& Ichikawa 1995). These derived values dominate the measured fluxes.  Since  Fukugita et al. (1995) does not provide the error estimate in the above filters,  we roughly estimated the error in the V, B, and U bands (which is $\sim$0.3\,mJy), by taking into account the error in R-band (0.26\,mJy). The estimated error in  B and U bands are comparable to, or larger than, the estimated host galaxy flux. Considering the rather large host galaxy magnitude (compared to the measured flux) and the high uncertainty,  the fluxes in these bands will not be considered in the (non-thermal) SED modeling in this paper. Indeed, in Sect.\,4 we will see that the {\it Swift}/UVOT data can be nicely interpreted as arising from the thermal emission of the elliptical host galaxy.

\subsection{{Fermi}/LAT}

\begin{figure}
   %\centering
\hspace{0cm}
\vspace{2.cm}
\includegraphics[width=0.45\textwidth]{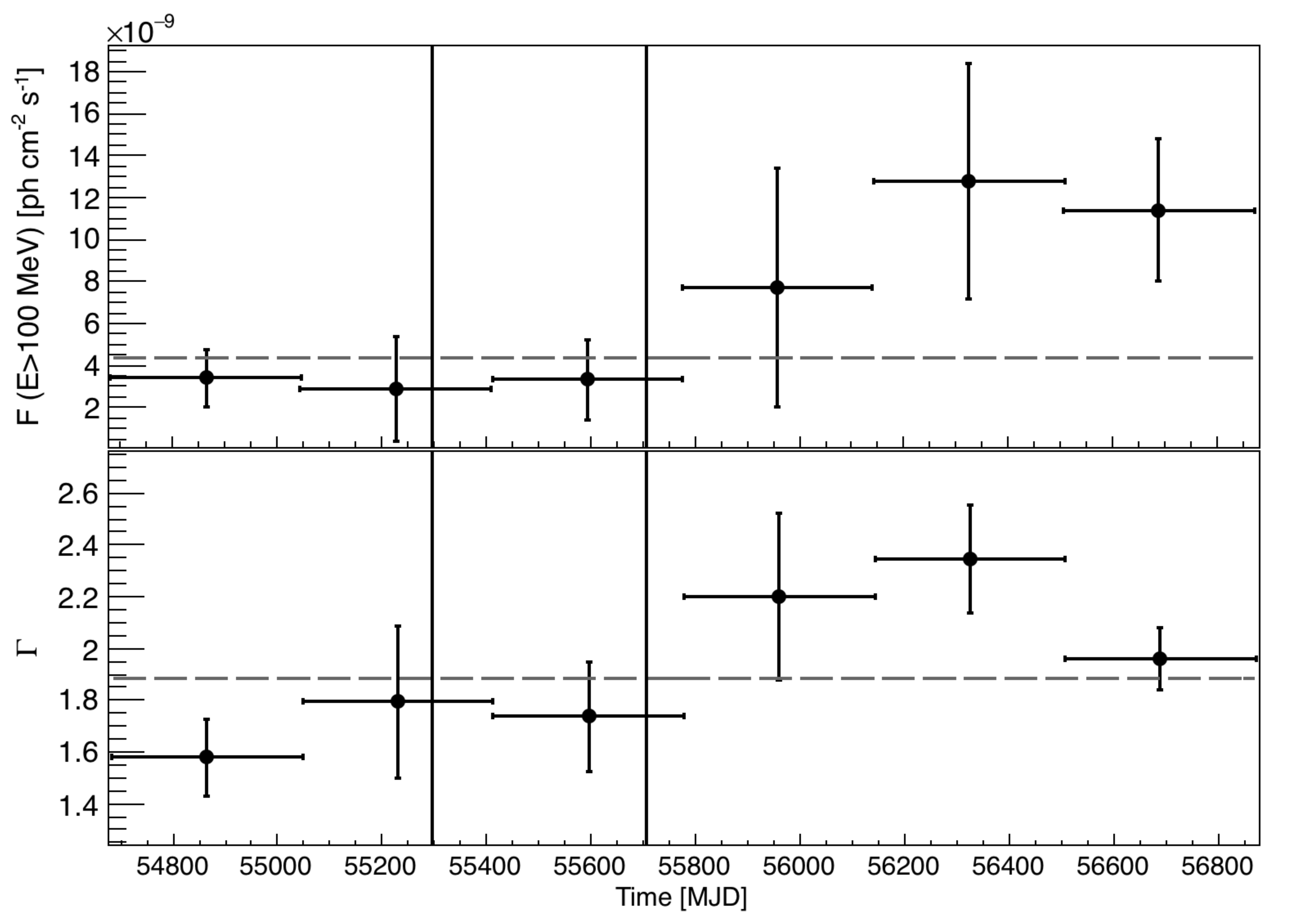} 
\vspace{-2.cm}
   \caption{1-year binning {\it Fermi}/LAT light curve for E$>$100\,MeV (top) and evolution of the spectral index as a 
function of the time (bottom). The grey dashed lines represent the mean flux and spectral index during 
the first 6.7 years of the mission, respectively. The period between the vertical bars denotes the MAGIC observation window.}
   \label{Fermi_LC}
\end{figure}

In Fig.~\ref{Fermi_LC} the {\it Fermi}/LAT fluxes and spectral indices of 1ES\,1741+196 are plotted as a function of time in bins of 1 year. Slight hints of variability, especially in the spectral index, do exist with $\chi^{2}_\nu = 8.1/5$ and 11.46/5 for, respectively, the flux and the spectral index, while fitting with a constant line.  Since the source was not bright enough, variability on shorter time scales  cannot be investigated. Also, no month-scale variability was found, compatible with previous claims in the 3FGL catalog (\citealp{fermi3} with a variability index of 38.3).

We have analysed HE {\it Fermi}/LAT data contemporaneous to the MAGIC observations. The LAT data were collected from 2010 April 10 (MJD 55296) to 2011 May 26 (MJD 55707). 
%A point like source positionally consistent with 1ES1741+196 was detected with  $ \sqrt{TS}=4.4 \sigma$ at E$>$100\,MeV. 
A point like source positionally consistent with 1ES\,1741+196 was detected with a  TS = 19.4 ($\sim 4.4\,{\mathrm{\sigma}}$).
The best-fit parameters for the model  result  in a spectral index of $\Gamma=1.6 \pm 0.1$  and an integral flux F(E$>$100\,MeV)=(2.0 $\pm$ 0.4)$\cdot 10^{-9} \mathrm{ph} \, \mathrm{cm}^{-2} \, \mathrm{s}^{-1}$.  The spectral index reported in the 3FGL  \citep{fermi3} is $1.8\pm0.1$, while the one reported in 1FHL \citep{1fhl} is  $2.1\pm0.5$. For comparison purposes,  a spectrum has also been produced for the whole data sample collected by the LAT from 2008 August 4 to 2015 April 7 (MJD 54682.7--57119.3) above 100\,MeV using the same procedure. %Both SEDs are shown in Fig.7.

\begin{figure}
%\centering
\vspace{0.cm}
\hspace{0.cm}
\includegraphics[width=0.5\textwidth]{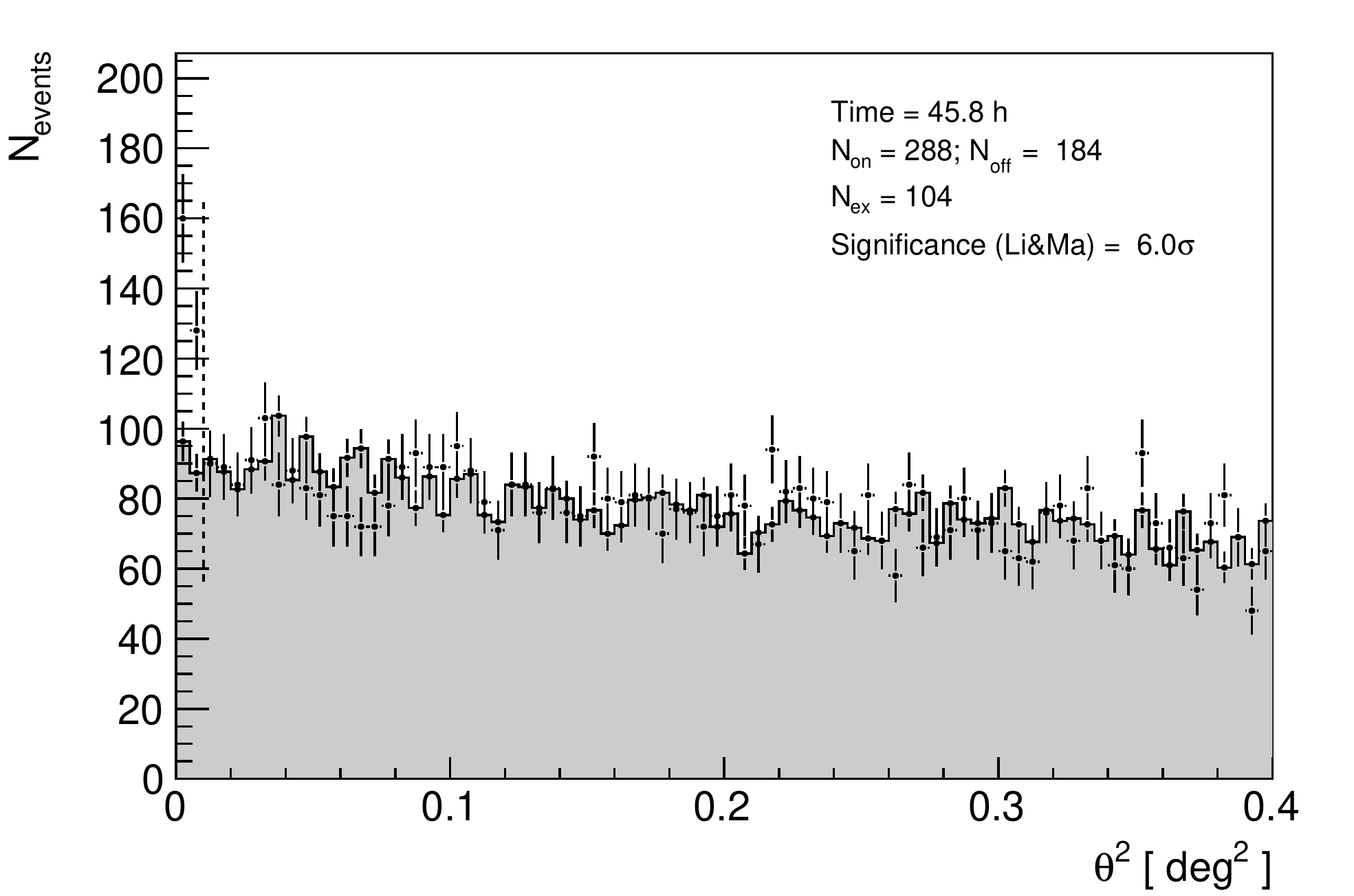}
\caption
{
$\theta^{2}$ distributions of the 1ES\,1741+196 signal and background estimation from 46\,hr 
of MAGIC stereo observations for $E > 250$\,GeV. The region between zero and 
the vertical dashed line (at $0.01$~$\mbox{degrees}^2$) represents the signal region.
}
\label{fig:theta2}
\end{figure}

\begin{figure}
\vspace{-3cm}
\hspace{-0.5cm}
\includegraphics[width=0.55\textwidth]{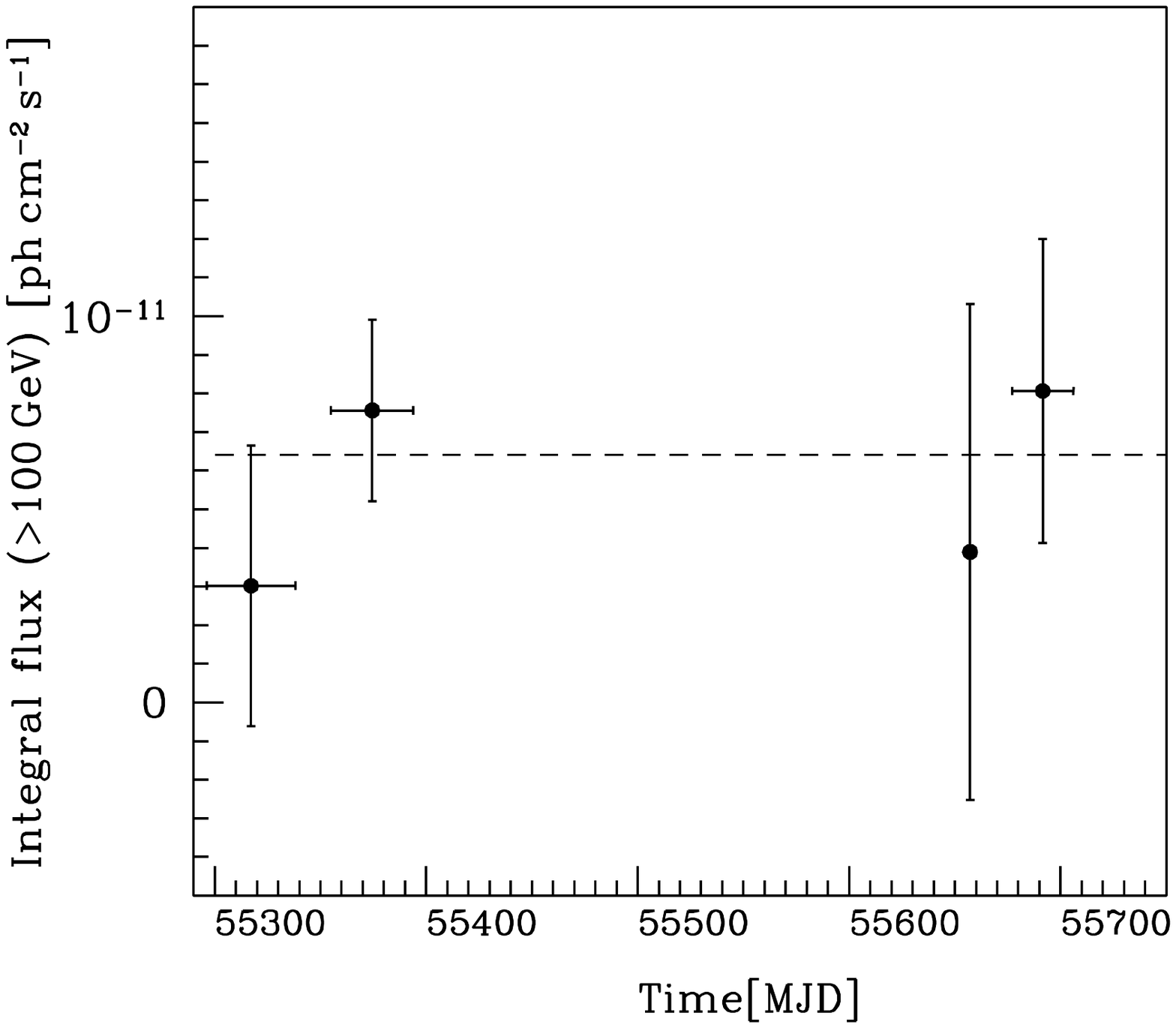}
\vspace{-4cm}
\caption{1ES\,1741+196 light curve during MAGIC observation period in a time bin of 28 days. The horizontal line represents the fit to the data assuming a constant flux (see text for details).}
\label{fig:lc_vhe}
\end{figure}

\subsection{MAGIC}

The $\gamma$-ray signal from the source is estimated from the so-called $\theta^2$ plot, after the application of energy-dependent cuts to events (including {\it hadronness}), and within a fiducial $\theta^2$ signal region. In order to evaluate the residual background of the observation, the $\theta^2$ distribution around a nominal background control region is also calculated. Fig.~\ref{fig:theta2} shows the $\theta^2$ distribution of the events.
We found an excess of $104\pm8$ events in the fiducial signal region with $\theta^2 < 0.01$~$\mbox{degrees}^2$, corresponding to a significance of $6.0\,\sigma$. 

 Fig.\,\ref{fig:lc_vhe} shows the light curve of the source with a time binning of 28 days, { considering the length of a moon-cycle, which determines the observational season of IACTs}. During the observation period no significant variability was detected. The light curve can be fitted with a constant flux hypothesis of $(6.4\pm1.7)\cdot10^{-12}$\,ph\,cm$^{-2}$s$^{-1}$ with a $\chi^{2}/{\mathrm{d.o.f.}}$=1.4/3. 

\begin{figure}%[htp]
\centering
\vspace{-1.5cm}
\includegraphics[width=0.5\textwidth]{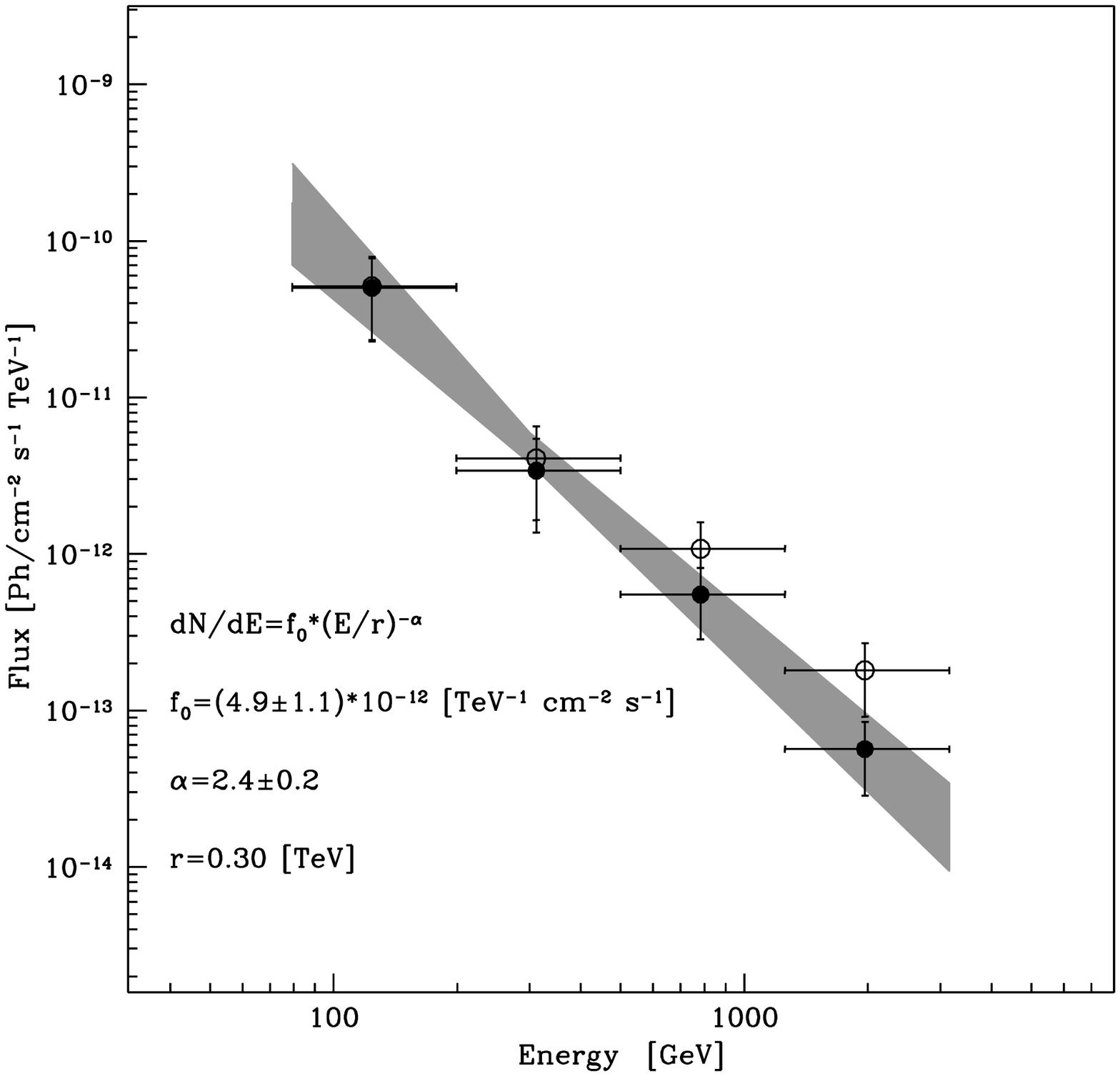}
\vspace{-2.5cm}
\caption{ 1ES\,1741+196 differential energy spectrum measured by MAGIC (filled circle), and EBL corrected spectrum using the Franceschini et al. (2008) model (empty circle). The fit function is shown as a grey bow-tie.}
\label{fig:spectrum}
\end{figure}

 Fig.\,\ref{fig:spectrum} shows the the spectral points of 1ES\,1741+196, which are obtained by the Bertero unfolding method (Albert et al. 2007). We also show the spectral points after correcting the Extragalactic Background Light (EBL) using the  Franceschini, Rodighiero \& Vaccari (2008) model. The bow-tie shows the power law fit obtained with the forward-folding method.
The spectrum in the range $\sim$80\,GeV$ <$ E $<$ 3\,TeV can be well described by a simple power law ($\chi^2/\mathrm {d.o.f.}=2.16/6$):

\begin{equation}
\mathrm{\frac{\mbox{d}N}{\mbox{d}E} = f_{0} \left(\frac{E}{\mathrm{0.3~TeV}}\right)^{-\alpha}},
\end{equation}
with a photon index of $\mathrm{\alpha = 2.4 \pm 0.2_{\mathrm{stat}}}\pm 0.2_{\mathrm{syst}}$,
and a normalization constant at 0.3\,TeV of $\mathrm{f_0} = (4.9 \pm 1.1_{\mathrm{stat}} \pm 0.9_{\mathrm{syst}}) \cdot 10^{-12}$ ph\,cm$^{-2}$\,s$^{-1}$\,TeV$^{-1}$. 
The systematic error  on the energy scale is 17\% \citep{aleksic+11}. 
The mean integrated flux above 100\,GeV is %$ \mathrm{F(> 100\,{\rm GeV}) = (6.4 \pm 1.7) \cdot 10^{-12}$ \,erg\,cm$^{-2}$s$^{-1}}$.
$\mathrm{F}(> 100\,{\rm GeV}) = (6.4 \pm 1.7_{\mathrm{stat}} \pm 2.6_{\mathrm{syst}}) \cdot 10^{-12}$\,ph\,cm$^{-2}$s$^{-1}$.
%

%
%This is the first measurement of the differential energy 
%spectrum of 1ES\,1741+196 at VHE\,$\gamma$-rays.
%

\section{Discussion}

\begin{figure}
\centering
%\hspace{-1.5cm}
%\vspace{-2.cm}
%\hspace{-3.cm}
\vspace{0.5cm}
\hspace{0.cm}
\includegraphics[width=9.cm]{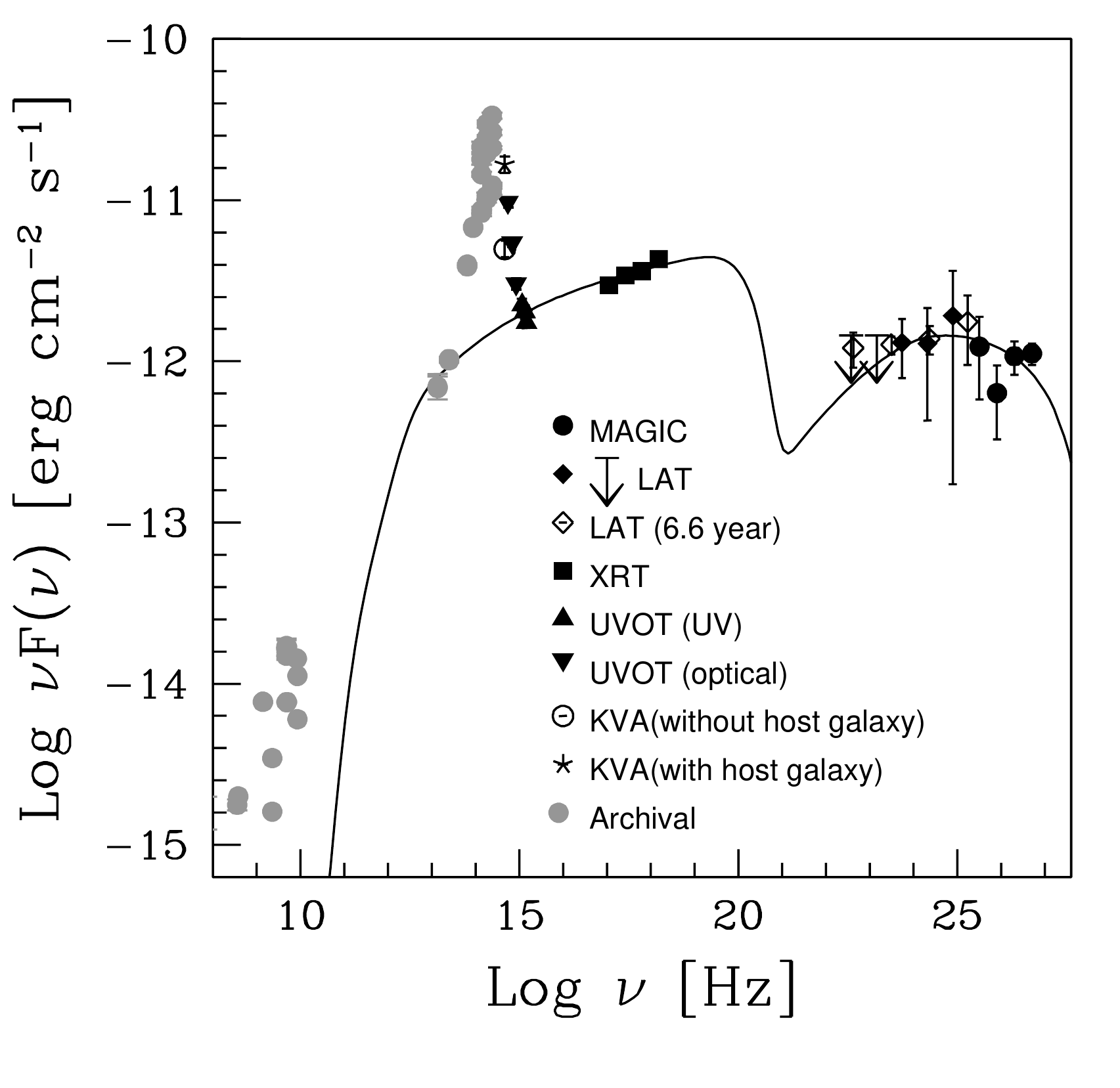}
\vspace{0cm}
\caption{The SED of 1ES\,1741+196 from eV to TeV energies. 
The UV (UVOT UVW1, UVM2, UVW2: filled triangle) fluxes attributed to the nuclear region of blazar are used for the SED fitting -- whereas the optical fluxes (KVA: empty circle; UVOT V, B, U: filled inverted triangle) are not. The UVOT UV fluxes are corrected for both the Galactic extinction and the host galaxy, while UVOT optical fluxes are corrected only for the former. The KVA flux with (empty circle) and without (star)  subtracting the host galaxy magnitude are also shown.
X-ray data (filled square) come from {\it Swift}/XRT, averaged over two distinct observations. Contemporaneous (filled diamond) and 6.6\,yr-integrated (empty diamond; for comparison) HE\,$\gamma$-ray data come from {\it Fermi}/LAT. 
MAGIC data (filled circle) are EBL corrected (Franceschini et al. 2008).  We also show the archival data (filled gray circle) for comparison.}
\label{fig:SED}
\end{figure}

\begin{figure}
\centering
\hspace{-1.5cm}
\vspace{-2.5cm}
\includegraphics[width=9.5cm]{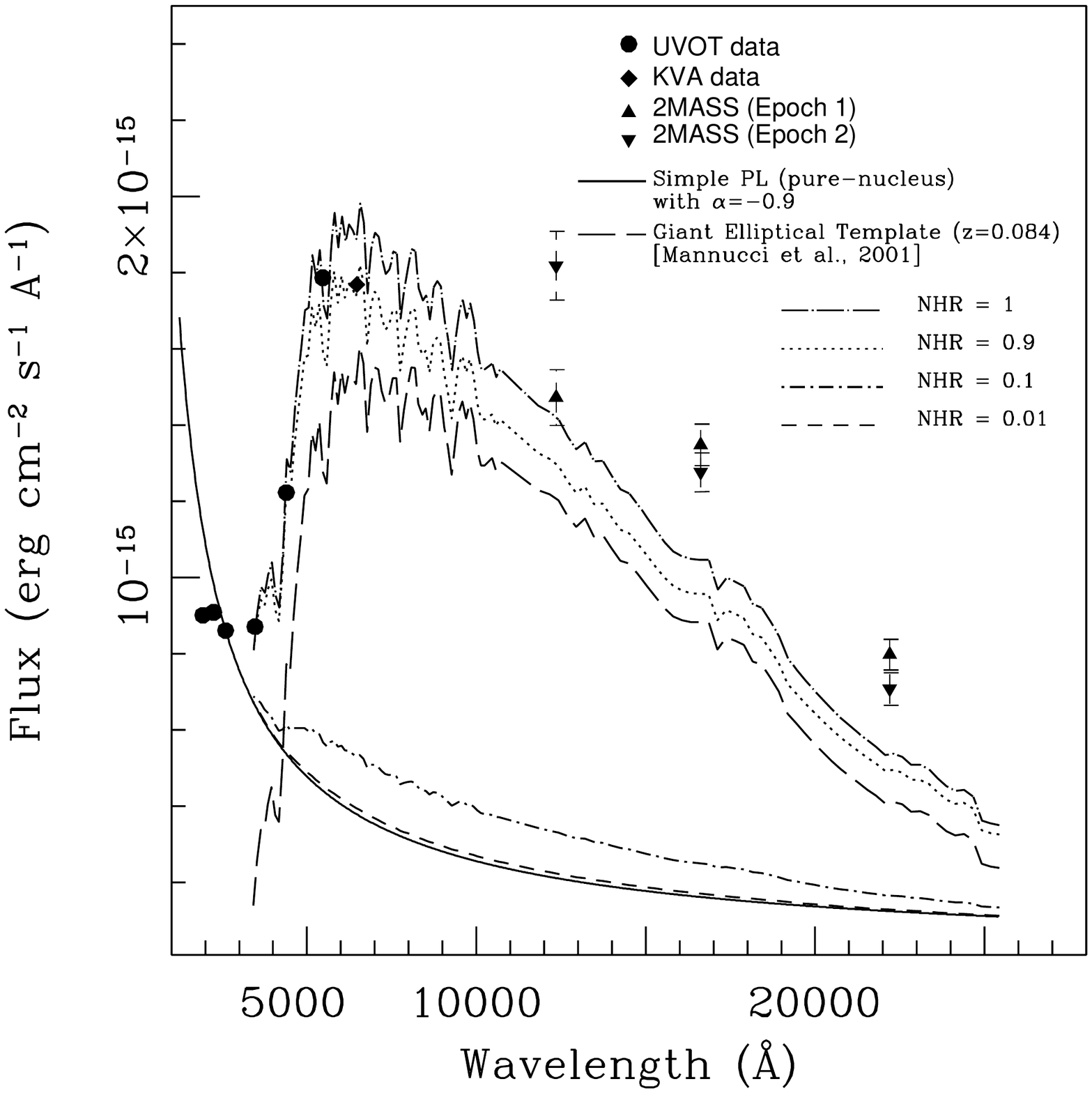}
\vspace{0cm}
\caption{1ES\,1741+196's spectral feature at optical/NIR frequencies, decomposed into blazar and host galaxy components. 
In the relevant spectral range of interest here, the blazar and the host galaxy are modelled as, respectively, a power law (with a spectral index of 0.9) and a giant-elliptical thermal template  (with $U-B=0.50$, $B-V=0.99$, $V-R = 0.59$, $V-I=1.2$2, $V-K=3.30$; see \citealp{mannucci+01}).
}
\label{fig:uvotsky}
\end{figure}

The emission of most BL Lac sources is understood in terms of  the one-zone SSC model, in which energetic electrons moving inside a magnetized relativistically-moving plasma blob emit via the synchrotron and inverse-Compton scattering mechanisms (e.g., Tavecchio et al. 1998). The electron spectrum is often  described as a smoothed broken power law.
\begin{equation}
N(\gamma)= \left\{\begin{matrix}
{\mathrm{K \gamma^{-n_{1}}}} \,\,\,\,\,\,\,\,\,\,\,\,\,\,\,\,\,\,;    {\mathrm{\gamma_{min} < \gamma < \gamma_{br}}}
\\ 
 {\mathrm{K \gamma_{br}^{n_{2}-n_{1} } \gamma^{-n_{2}}}} \,\,; {\mathrm{\gamma_{br} < \gamma < \gamma_{max}}}
\end{matrix}\right\}
\end{equation}
where $\mathrm{\gamma_{min}}$, $\mathrm{\gamma_{\rm br}}$, and $\mathrm{\gamma_{max}}$ are the lowest, break, and highest Lorentz factors, K is the normalization constant, and $\mathrm{n_1}$ and $\mathrm{n_2}$ are, respectively, the slopes below and above the break. The relativistic boosting is encoded in the Doppler factor $\mathrm{\delta = [\Gamma 
({1-{\mathrm{\dfrac{v}{c}}} {\rm cos}\,\theta})]^{-1}}$,  where $\Gamma$ is the bulk Lorentz factor, and   $\theta$ is the viewing angle.

%As mentioned before, 1ES\,1741+196 is peculiar BL Lac, since 
%it is in a triplet of interacting galaxies. The nearby 
%companion galaxies - an Sa-type galaxy and a spiral galaxy - 
%are at a projected distance of 3.3" and 12" respectively. The 
%UVOT image of the region is shown in Fig.\,6: as it can be 
%seen in the figure, the two companion galaxies are brightest 
%in the V, B, and U bands, and least bright in the UW band. 
%The signal region of UVOT (a circle of radius 5" around the 
%position of 1ES\,1741+196) may include a contribution from all 
%the three galaxies (Sa, spiral, host elliptical host). It was 
%not possible to 
%subtract the UVOT flux points in V, B and U band from the 
%combined flux from the three galaxies (see section\,3.2). The 
%optical flux point from the KVA, on the other hand is at a 
%higher flux level, compared to the trend obtained from the 
%spectral points of {\it Swift}/XRT and the {\it Swift}/UVOT 
%in the UV band. Considering these differences, the optical 
%points are ignored for the emission modeling of the source.} 

In Fig.6 we plot the broad-band SED of 1ES\,1741+196 using the multiwavelength data described in Sect.\,3. The VHE data are corrected for the EBL using the Franceschini et al. (2008) model. The {\it Fermi}/LAT spectrum (HE\,$\gamma$-ray band),  which is contemporaneous with the MAGIC observation is shown as filled diamonds, while the {\it Fermi} spectrum for the total time period (empty diamonds) is not used for the SED modeling. The {\it Swift}/XRT X-ray spectrum (averaged over the nights of 30 Jul 2010 and 21 Jan 2011) is also depicted. The optical KVA point,  that represents the blazar's non-thermal optical flux (free from the host galaxy emission), was computed convolving a point-like source with a typical KVA seeing of 2 arcsec. 

The SED of 1ES\,1741+196 does not look very different from the SEDs of other BL\,Lacs (\citealp{tav+10}). However, it shows an unusual conspicuous feature at optical/NIR frequencies, that appears as a branch taking off from the familiar non-thermal SED. We assume this feature to be the spectral signature of the elliptical galaxy hosting the blazar. In Fig.7 we overlay the optical/NIR data --{\it Swift}/UVOT and KVA -- together with the non-simultaneous data from the 2 Micron All Sky Survey (2MASS) obtained on 1999 June 19\footnote{http://www.ipac.caltech.edu/2mass/}, onto the giant elliptical template of Mannucci et al. (2011), placed at the appropriate distance: the good agreement below 10000\,${\rm \AA}$ supports our assumption. The difference between the model and the 2MASS flux  could be due to the contribution of the two nearby galaxies in the triplet which was not taken into account in this modeling.

We fitted the {\it non-thermal} SED using the method described in \citet{mankuzhiyil+11}, i.e., assuming a one-zone SSC emission model (Tavecchio et al. 1998). The   SSC model parameters obtained are given in Table~\ref{table:2}. Fitting an SSC model to the observed SED data returns parameters that are typically found for HBL (e.g., \citealp{tav+10}, \citealp{mankuzhiyil+11}, \citealp{mankuzhiyil+12}),  except for the comparatively lower Doppler factor, and the minor difference between the slopes below and above the $\gamma_{break}$ of the electron energy distribution (EED).  We note  that the experimental  constraints are relatively limited,  hence the  SSC  parameter  combination may  not  be  unique.  Alternative combinations of parameters could also provide an acceptable fit to the data. 

Our results are in overall agreement with those recently reported by the VERITAS collaboration (\citealp{veritas}), pertaining to observations made after ours. The similar results should not be surprising given that  the source is consistent with the hypothesis of no variability  during the last $\sim$6 years.

 It is interesting to note that the infrared region of the archival data (which were not considered for the SED fit) nicely match with the SED. However, the KVA (host galaxy subtracted) flux does not agree well with the non-thermal SED, while the KVA flux with the host galaxy shows a rough agreement between the non-thermal infrared-optical flux points. The lack of other  host-galaxy-subtracted data in the optical region  prevents us from testing any other emission model (see for example, the helical jet model of Villata \& Raiteri 1999, that was applied in Ahnen et al. 2016).

 The radio emission is plausibly from an extended region, hence does not agree with the model  below the frequency $10^{11}$\,Hz. As a demonstration, we calculate the typical Lorentz factor ($\gamma_{\mathrm{R}}$) of electrons that are responsible for the radio emission at $\sim$ 10\,GHz in a magnetic field $3.9\times10^{-2}$\,G and Doppler factor 14 (from the Table\,1). The calculated value  turns out to be  $\gamma_{\mathrm{R}}\sim$100.  The cooling time scale of such an electron will be $t_{\mathrm{c}}\sim 1.\times10^{10}$\,s. Assuming the electron diffuses through the jet medium at the speed of light, the extent of the radio emitting region will be $\sim$ 100\,pc, which is far beyond the blob radius. This suggests that the radio is emitted from an extended region, hence may differ from the model.

%\begin{figure}
%\centering
%\includegraphics[width=8.5cm]{image_uvot_all_con.jpeg}
%\caption{{ (To be replaced with a better a pic.)}
%The region of 1ES\,1741+196 as viewed by the {\it Swift}/UVOT. The bands (in the order from the top left to bottom right) are UW1, UW2, UM2, U, B and V. The position of 1ES\,1741+196 is marked as a red square. The signal area is marked as a green circle of 5'' radius. The Sa-type galaxy stays within the circle. The spiral galaxy is at 12'' from the BL\,Lac. 
%}
%\label{fig:uvotsky}
%\end{figure}

%\begin{figure}%[htp]
%\centering
%\vspace{-0.cm}
%\includegraphics[width=0.5\textwidth]{1741_spec_therm}
%\vspace{0cm}
%\caption{1ES\,1741+196's spectral feature at optical/NIR frequencies, decomposed into blazar and host galaxy components. In the relevant spectral range of interest here, the blazar and the host galaxy are modelled as, respectively, a power law (with a spectral index of 1.7) and a giant-elliptical thermal template (with (with $U-B=0.50$, $B-V=0.99$, $V-R = 0.59$, $V-I=1.2$2, $V-K=3.30$; see \citealp{mannucci+01}.}
%\label{fig:spectrum}
%\end{figure}

%

%
%
\begin{table}%[htp]
\caption{Model parameters used for fitting the SED in Fig.6.}
\label{table:2}
\centering
\setlength{\tabcolsep}{0.40em}
%\begin{tabular}{l l l l l l l r l l l }
\begin{tabular}{l l l l l l l l l}

\hline

\tiny $\mathrm{\gamma_{min}}$ &\tiny $\mathrm{\gamma_{br}}$ &\tiny $\mathrm{\gamma_{max}}$ &\tiny $\mathrm{n_{1}}$ &\tiny $\mathrm{n_{2}}$ &\tiny $\mathrm{K\,[cm^{-3}]}$ &\tiny $\mathrm{B\,[G]}$ &\tiny $\mathrm{R\,[cm]}$&\tiny $\mathrm{\delta}$ \\
\hline
    \tiny $1\cdot 10^{3}$ & \tiny 5.1$\cdot 10^{3}$ & \tiny 9.4$\cdot 10^{6}$ &\tiny 2.2 &\tiny 2.9 &\tiny 2.3$\cdot 10^{4}$ & \tiny 3.9$\cdot 10^{-2}$&\tiny 2.0$\cdot 10^{16}$& \tiny14.0
\end{tabular}
\end{table}

  We note that the   Doppler factor from the SSC fit is well above the Doppler factor ($>4$) calculated from the jet-counterjet radio brightness ratio \citep{radiomap}. This is a common dispute in blazars, where the Doppler factor from the SSC fit falls mostly in the range of 10-50 (\citealp{tav+10}), while it is a few from the radio brightness studies (\citealp{radiomap2}). This may be because the Doppler factor that we estimate through the SED modeling belongs to the blazar zone. The size of the emission region derived from the SSC fit is $2\times 10^{16}$ cm. If we assume a conical jet of opening angle of 1$^\circ$, then the blazar emission region is located at $\sim 1\times 10^{18}$cm from the central engine.  This distance corresponds to an angular separation of $\sim$0.1 milli  arcsec (at a redshift z=0.084), which is beyond the resolution of radio telescopes. However, the jet-counterjet brightness ratio is estimated from the extended region of the jet.

Perhaps related to 1ES\,1741+196's host galaxy being visible, the equivalent isotropic luminosity, estimated from the peak fluxes and the corresponding frequencies  of the synchrotron and SSC components of the SED of this source, $\sim 8.2 \times 10^{43}$ erg\,s$^{-1}$, is among the lowest among TeV blazars. This is at least partly due to its Doppler factor, $\delta \simeq 14$, being lower by a factor of $\sim$2 than typical values found in TeV blazars -- maybe owing to misalignment -- as $L \propto \delta^4$, this source may indeed appear underluminous by a factor of $\sim$20. 

  The ${\mathrm{\gamma_{break}}}$ of the electron energy distribution (EED) which lies near the ${\mathrm{\gamma_{min}}}$, and the minor difference in the EED  slopes  below and above the  ${\mathrm{\gamma_{break}}}$ (2.2 and 2.9 respectively) are unusual compared to the EED parameters of other BL\,Lacs (see for example, \citealp{tav+10}, \citealp{mankuzhiyil+11}, \citealp{mankuzhiyil+11}). The reader may also note that the VERITAS collaboration reported a simple power law EED (instead of a broken power law that we use) of slope 2.2 to explain the emission process of this source. 
%{\bf We have also attempted to fit the SED by assuming a simple power law EED. However, the obtained SED could not well replicate the optical and VHE band. A small difference in the EED slopes was reported by Mankuzhiyil et al. (2012) to explain the short-term flaring of Mrk\,501. In contrast, in our case, the source is consistent with the hypothesis of no variability.} 
 We have also attempted to fit the SED using a simple power-law EED, and found that the model does not reproduce well the UV band (connected to the X-ray spectrum) and the flatness of the measured high-energy (IC) peak. Therefore, one needs a double power-law EED with an internal break with a relatively small spectral change ($\Delta_{n}$ =0.7) to describe well the measured broadband SED reported in this study. The origin of these internal breaks in the EED, presumedly related to the acceleration process, may be due to variations in the global field orientation or turbulence levels sampled by particles of different energy. The need for this kind of internal breaks in the EED have been reported in the literature for several sources in order to better describe the spectral measurements. Examples of those are the ones reported for 3C\,454 \citep{3c454}, AO\,0235+164 \citep{0235}, Mrk\,421 (\citealp{abdo421}, \citealp{aleksic421}) and Mrk\,501 (\citealp{abdo501} and \citealp{mankuzhiyil+12} during short flares).
A detailed study on the emission process will be addressed in a more detailed paper.

 To the best of our knowledge, 1ES\,1741+196 is the first blazar with known SED hosted in a triplet of interacting galaxies.  It is interesting to note that, even though a tidal tail is observed to emanate from the host galaxy (\citealp{heidt1999}) -- suggesting mass loss from the galaxy due to tidal forces within the triplet -- the SSC emission parameters of 1ES\,1741+196 do not substantially deviate from typical values (except the Doppler factor and the slope of EED) observed in other BL\,Lacs. 

%This should not be suprising, because the size scales affected by tidal stripping and those relevant for nuclear BH accretion are not coupled. 

\section{Summary}

We reported the first detection (by MAGIC) of VHE\,$\gamma$-rays from the BL\,Lac source 1ES\,1741+196. From the 2010-2011 MAGIC data the source is clearly detected at $6.0\,\sigma$ significance level. The multi-frequency data used in this paper  suggest that 1ES\,1741+196 shows a behaviour typical of HBL sources, with a slightly different EED and a lower Doppler factor.
 A notable peculiarity of the SED of 1ES\,1741+196 is that it shows the host galaxy's spectral signature, a thermal feature at optical/NIR frequencies that we show is compatible with the spectrum of a giant elliptical. The coincidental relatively low luminosity of 1ES\,1741+196 may stem from the jet's relatively low Doppler factor.

\section*{acknowledgements}

We would like to thank
the Instituto de Astrof\'{\i}sica de Canarias
for the excellent working conditions
at the Observatorio del Roque de los Muchachos in La Palma.
The financial support of the German BMBF and MPG,
the Italian INFN and INAF,
the Swiss National Fund SNF,
the ERDF under the Spanish MINECO
(FPA2015-69818-P, FPA2012-36668, FPA2015-68278-P,
FPA2015-69210-C6-2-R, FPA2015-69210-C6-4-R,
FPA2015-69210-C6-6-R, AYA2013-47447-C3-1-P,
AYA2015-71042-P, ESP2015-71662-C2-2-P, CSD2009-00064),
and the Japanese JSPS and MEXT
is gratefully acknowledged.
This work was also supported
by the Spanish Centro de Excelencia ``Severo Ochoa''
SEV-2012-0234 and SEV-2015-0548,
and Unidad de Excelencia ``Mar\'{\i}a de Maeztu'' MDM-2014-0369,
by grant 268740 of the Academy of Finland,
by the Croatian Science Foundation (HrZZ) Project 09/176
and the University of Rijeka Project 13.12.1.3.02,
by the DFG Collaborative Research Centers SFB823/C4 and SFB876/C3,
and by the Polish MNiSzW grant 745/N-HESS-MAGIC/2010/0.
The \textit{Fermi} LAT Collaboration acknowledges generous ongoing support
from a number of agencies and institutes that have supported both the
development and the operation of the LAT as well as scientific data analysis.
These include the National Aeronautics and Space Administration and the
Department of Energy in the United States, the Commissariat \`a l'Energie Atomique
and the Centre National de la Recherche Scientifique / Institut National de Physique
Nucl\'eaire et de Physique des Particules in France, the Agenzia Spaziale Italiana
and the Istituto Nazionale di Fisica Nucleare in Italy, the Ministry of Education,
Culture, Sports, Science and Technology (MEXT), High Energy Accelerator Research
Organization (KEK) and Japan Aerospace Exploration Agency (JAXA) in Japan, and
the K.~A.~Wallenberg Foundation, the Swedish Research Council and the
Swedish National Space Board in Sweden. Additional support for science analysis during the operations phase is gratefully acknowledged from the Istituto Nazionale di Astrofisica in Italy and the Centre National d'\'Etudes Spatiales in France. This research was supported ´ 572 by an appointment to the NASA Postdoctoral Program at the Goddard Space Flight Center, administered by Universities Space Re- 574 search Association through a contract with NASA.
F. K. acknowledges funding from the European Union’s Horizon 2020 research and innovation programme under grant agreement No 653477.
%\end{acknowledgements}

\noindent
$^{1}$ {ETH Zurich, CH-8093 Zurich, Switzerland} \\
$^{2}$ {Universit\`a di Udine, and INFN Trieste, I-33100 Udine, Italy} \\
$^{3}$ {INAF National Institute for Astrophysics, I-00136 Rome, Italy} \\
$^{4}$ {Universit\`a  di Siena, and INFN Pisa, I-53100 Siena, Italy} \\
$^{5}$ {Universit\`a di Padova and INFN, I-35131 Padova, Italy} \\
$^{6}$ {Croatian MAGIC Consortium, Rudjer Boskovic Institute, University of Rijeka, University of Split and University of Zagreb, Croatia} \\
$^{7}$ {Saha Institute of Nuclear Physics, 1/AF Bidhannagar, Salt Lake, Sector-1, Kolkata 700064, India} \\
$^{8}$ {Max-Planck-Institut f\"ur Physik, D-80805 M\"unchen, Germany} \\
$^{9}$ {Universidad Complutense, E-28040 Madrid, Spain} \\
$^{10}$ {Inst. de Astrof\'isica de Canarias, E-38200 La Laguna, Tenerife, Spain; Universidad de La Laguna, Dpto. Astrof\'isica, E-38206 La Laguna, Tenerife, Spain} \\
$^{11}$ {University of \L\'od\'z, PL-90236 Lodz, Poland} \\
$^{12}$ {Deutsches Elektronen-Synchrotron (DESY), D-15738 Zeuthen, Germany} \\
$^{13}$ {Institut de Fisica d'Altes Energies (IFAE), The Barcelona Institute of Science and Technology, Campus UAB, 08193 Bellaterra (Barcelona), Spain} \\
$^{14}$ {Universit\"at W\"urzburg, D-97074 W\"urzburg, Germany} \\
$^{15}$ {Institute for Space Sciences (CSIC/IEEC), E-08193 Barcelona, Spain} \\
$^{16}$ {Technische Universit\"at Dortmund, D-44221 Dortmund, Germany} \\
$^{17}$ {Finnish MAGIC Consortium, Tuorla Observatory, University of Turku and Astronomy Division, University of Oulu, Finland} \\
$^{18}$ {Unitat de F\'isica de les Radiacions, Departament de F\'isica, and CERES-IEEC, Universitat Aut\`onoma de Barcelona, E-08193 Bellaterra, Spain} \\
$^{19}$ {Universitat de Barcelona, ICC, IEEC-UB, E-08028 Barcelona, Spain} \\
$^{20}$ {Japanese MAGIC Consortium, ICRR, The University of Tokyo, Department of Physics and Hakubi Center, Kyoto University, Tokai University, The University of Tokushima, KEK, Japan} \\
$^{21}$ {Inst. for Nucl. Research and Nucl. Energy, BG-1784 Sofia, Bulgaria} \\
$^{22}$ {Universit\`a di Pisa, and INFN Pisa, I-56126 Pisa, Italy} \\
$^{23}$ {ICREA and Institute for Space Sciences (CSIC/IEEC), E-08193 Barcelona, Spain} \\
$^{24}$ {also at the Department of Physics of Kyoto University, Japan}\\
$^{25}$ {now at Centro Brasileiro de Pesquisas F\'isicas (CBPF/MCTI), R. Dr. Xavier Sigaud, 150 - Urca, Rio de Janeiro - RJ, 22290-180, Brazil}\\
$^{26}${now at NASA Goddard Space Flight Center, Greenbelt, MD 20771, USA and Department of Physics and Department of Astronomy, University of Maryland, College Park, MD 20742, USA}\\
$^{27}${Humboldt University of Berlin, Institut f\"ur Physik Newtonstr. 15, 12489 Berlin Germany}\\
$^{28}${also at University of Trieste}\\
$^{29}${now at Ecole polytechnique f\'ed\'erale de Lausanne (EPFL), Lausanne, Switzerland}\\
$^{30}${now at Max-Planck-Institut fur Kernphysik, P.O. Box 103980, D 69029 Heidelberg, Germany}\\
$^{31}${now at Astrophysical Sciences Division, BARC, Mumbai, India}\\
$^{32}${also at Japanese MAGIC Consortium}\\
$^{33}${now at Finnish Centre for Astronomy with ESO (FINCA), Turku, Finland}\\
$^{34}${also at INAF-Trieste and Dept. of Physics \& Astronomy, University of Bologna}\\
$^{35}${also at ISDC - Science Data Center for Astrophysics, 1290, Versoix (Geneva)}\\
$^{36}${now at IPNS, High Energy Accelerator Research Organization (KEK)
1-1 Oho, Tsukuba, Ibaraki 305-0801, Japan}\\
$^{37}${GRAPPA \& Anton Pannekoek Institute for Astronomy, University of Amsterdam,  Science Park 904, 1098 XH Amsterdam, The Netherlands}\\
$^{38}${ASI Science Data Center, Via del Politecnico snc I-00133, Roma, Italy}\\
$^{39}${NASA Goddard Space Flight Center, Greenbelt, MD 20771, USA}\\

\end{document}